# Communication in Agile Software Development - A Mapping Study

Suddhasvatta Das[1] and Kevin Gary[1]

[1] Arizona State University, Tempe AZ 85287, USA

{sdas76,kgary}@asu.edu

**Abstract:** *Software industry is a fast-moving industry and to keep up with this pace the development process also needs to be fast and efficient and Agile software development (ASD) is the answer to this problem. Even though ASD has been in there for over two decades there are still multiple unknown questions tied to ASD that need to be addressed. In this study we are going to address one of the most critical factors of ASD i.e. Communication. We conducted a review of 14 studies and found the areas under ASD communication that the community is interested in as well as research gaps.*

**Keywords:** Agile, Communication, Mapping Study

**Introduction**

Communication and coordination are keys to the success of any project and especially software projects. As agile methods ensure light, fast and quick software solutions [1] it's clear that agile is the method that will be followed in the near future. Agile values and principles work for small, efficient, collocated teams that aim to deliver rapid values with small products and collaborating with customers. Continuous feedback from the customer is also an important aspect that develops shared understanding rather than just requirements that are predetermined [2].

The advent of agile methods and lean engineering principles over the past decade [3] changed the way software practitioners deal with complex tasks and volatility. Practitioners now employ more collaborative approaches [4] to obtain continuous feedback in the spirit of empirical process control [5]. Approaches like Planning Poker, Pair Programming, User Story Mapping [6] are some known examples of collaborative ways of working. These methods have also gained popularity in large-scale settings as well. However, there are challenges when multiple teams need to collaborate and coordinate in a hierarchical setup [7]. Mathieu et al. [8] defined this as a multiteam system (MTS).

According to [9] agile methods have impacts on the organization structure, culture and roles, contract negotiation and human resource. It also changes project practice and social interaction. The idea of agile methods was to deal with uncertainty and change and remove traditional coordination ways such forward planning, extensive documentation, strictly following a predefined process and specific roles [9]. Agile supports strong face to face communication and some simple practices and this proved to be successful for many projects. Therefore, there must have been an effective communication and coordination mechanism which has not been fully understood and explored [9].

This serves as a motivation for us, and we explore more into this area and provide the community with new knowledge that will help us to move forward. There has been multiple research investigating the communication and coordination aspect of agile methods. However, we could not identify any mapping studies that present the current state of research in this area and the research gaps in agile communication. Keeping this in mind this study aims to answer the following questions.

- *RQ1: What were the goals of previous studies that explored the area of agile communication?*
- *RQ2: What research gaps are there in the area of agile communication?*

**Background**

Agile software development (ASD) has been getting popular over the past couple of decades. A significant amount of research has been performed in this area, especially the area of communication. This section summarizes the relevant literature that have been included in this study.

The systematic mapping study by Ahmad, M. O., Lenarduzzi, V., Oivo, M., & Taibi [S1] states that both synchronous and asynchronous channels of communication are used in distributed and co-located ASD. The study also states that if face-to-face or synchronous communication is not possible then the use of asynchronous or email increases. This enhances useful and conclusive responses. The authors claim that daily standup meetings used with different communication tools also helps in synchronous communication. The concept of *one team attitude* is also stated in this article. Authors state that scrum planning meetings helps to reduce the geographical distance. Nonverbal communication such as body language, hand gestures also play an important role. The authors state that management needs to spend time and money on exchanging team members from different sites to understand the culture and ways of communication. Lastly, the paper points out that challenges in communication are not only limited to the communication media but include other parameters such as fixed requirements, process-oriented control, and customers with no decisive power. These challenges can be avoided by following strategies such as defining customer roles and having decisive power for the project's functionality and scope.

Alzoubi, Y. I., & Gill, A. Q [S2] present a Systematic Literature Review on communication challenges in agile GSD (Global Software Development). The study reviewed 22 papers and found 7 major classifications of challenges in agile GSD. The results show that people and distance difference were the top challenges reported. The other main challenge was architectural issues and role of the architect. The study also reports that using cloud-based communication media (Skype, Chatter, Yammer) can make communication in agile GSD better. Team gathering at the start of a project is also important to reduce cultural differences and increase collaboration. Visiting different locations, mandatory presentations from all team members and documentation are also some of the best practices to have a successful agile project.

Hummel, M., Rosenkranz, C., & Holten, R [S14] also presents a Systematic Literature review on the roles of communication in agile. Results of the study states that the knowledge on the role and impact of communication in agile software development (ASD) is not precise due to contradiction, scattered and inconclusive. Authors point out that the literature has a very broad view of communication in ASD, and prior studies have also suggested methods to improve communication but only a few of those considered the 'black-box' aspect of communication, social interactions and behavior of teams. This research also shows the lack of studies comparing communications among traditional software development and ASD.

**Research Methodology**

This section presents the study selection process and data extraction. The search process was conducted by the primary author and reviewed by the second author**.** The steps of the research process are shown in **Fig. 1** and summarized as follows:

1. Digital libraries from IEEE (169), ACM (22,513) and Google Scholar (268,000) were searched for conference, journal, and workshop articles.
2. For IEEE all the titles and abstracts were scanned, and 7 relevant articles were included in the next step. ACM and Google Scholar returned a huge number of papers with the same search strings. However, we identified that after selecting sort by relevance results page 3 (50 entries per page) of ACM and page 6 (10 entries per page) of Google Scholar the articles were not relevant to us and we did not look any further. We scanned all the titles and abstracts of all these articles and included 7 papers from ACM and 19 papers from Google Scholar.

3. Next, we removed 2 duplicates from the Google Scholar list of 17 papers.
4. The set of papers (31) was evaluated with respect to 3 quality attributes: *i) Studies before the year 2013 were excluded ii) Studies that had unclear evidence were excluded iii) Studies published at unknown venues were excluded.*
5. The final set of papers (14) in Table 1 was determined by a full-text manual review of the studies from step 4.

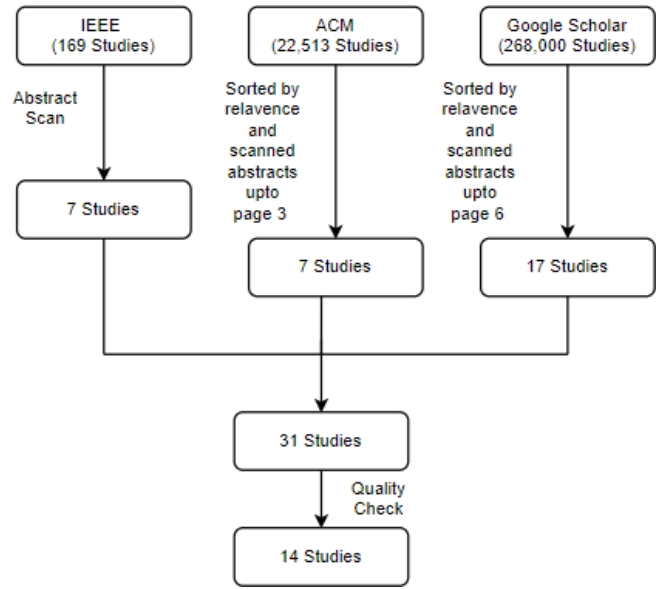

**Fig 1.** Search process to identify papers to include in this study.

**Table 1:** Final set of papers included in the study.

| Study | Venue | Year | Type | Citations | Research Goals/Focus |
|---|---|---|---|---|---|
| S1 | Conference | 2018 | Mapping Study | 7 | Role of communication in agile software development |
| S2 | Conference | 2016 | Survey | 10 | Impact and limitations of communication channels on virtual teams |
| S3 | Conference | 2019 | Case Study | 2 | Impact of communication patterns on quality and performance |
| S4 | Conference | 2016 | Case Study | 9 | Impact of SCRUM+KANBAN on geographically distributed teams |
| S5 | Conference | 2019 | Case Study | 5 | Communication benefits and impact on code refactoring, design and bug fix |
| S6 | Symposium | 2020 | Case Study | 0 | Communication challenges in agile |
| S7 | Conference | 2019 | Experience Report | 7 | Experience report of interpersonal communication of distributed agile teams |
| S8 | Conference | 2019 | Systematic Review + Semi Structured Interviews | 2 | How customer communication problems are solved in agile requirements engineering |
| S9 | Workshop | 2013 | Survey | 12 | Impact of communication on trust |
| S10 | Conference | 2013 | Case Study | 36 | Communication factors for speed and reuse in large scale agile development |
| S11 | Conference | 2014 | Systematic Review | 41 | Challenges of communication in ASD |
| S12 | Conference | 2014 | Case Study | 10 | Impact of infrastructure on communication |
| S13 | Journal | 2020 | Case Study | 5 | Agile communication issues in startups |
| S14 | Journal | 2013 | Systematic Review | 109 | Role of communication in agile software development |

We extracted information from the 14 studies identified in Table 1. Our analysis was focused on the overall research goals of each of these papers, not the specific research questions addressed by these studies. These themes are given in the rightmost column of Table 1.

We acknowledge limitations to this process. First, the area of communication in ASD has been explored by researchers since the mid-2000s and in this study we included studies from 2013 onwards as we sought to review more recent issues as agile adoption at scale continues to grow. Our selection of 2013 was based on our experience with the literature in the scaled agile space but could be considered arbitrary. Second, the digital libraries ACM and Google Scholar returned more than 200,000 studies based on our search string. Therefore, we sorted all the results by relevance and included relevant studies till 3rd and 6th page of ACM and Google Scholar, respectively as that is when we saw a continued absence of relevant papers. However, it is possible we might have missed some studies. Finally, this study was performed by a PhD student and a single secondary reviewer and a third reviewer might have improved the study.

**Discussion**

In this section we revisit our Research Questions.

*RQ1: What were the goals of previous studies that explored the area of agile communication?*

From Table 1 we found out that 4 (S2, S6, S11, S13) out of 14 studies focused on the challenges or issues related to communication in agile software development (ASD). 2 (S1, S14) of 4 studies discussed the role of communication ASD. 6 (S2, S3, S4, S5, S9, S12) out 14 studies had research goals directed towards the impact of communication in ASD. There was only one study that focused on interpersonal communication in distributed ASD teams (S7), how to resolve customer communication issues for requirements engineering (S8) and communication factors that speed up the development in large scale ASD respectively (S10). Figure 2 below shows the count of studies in each area related to the communication in ASD.

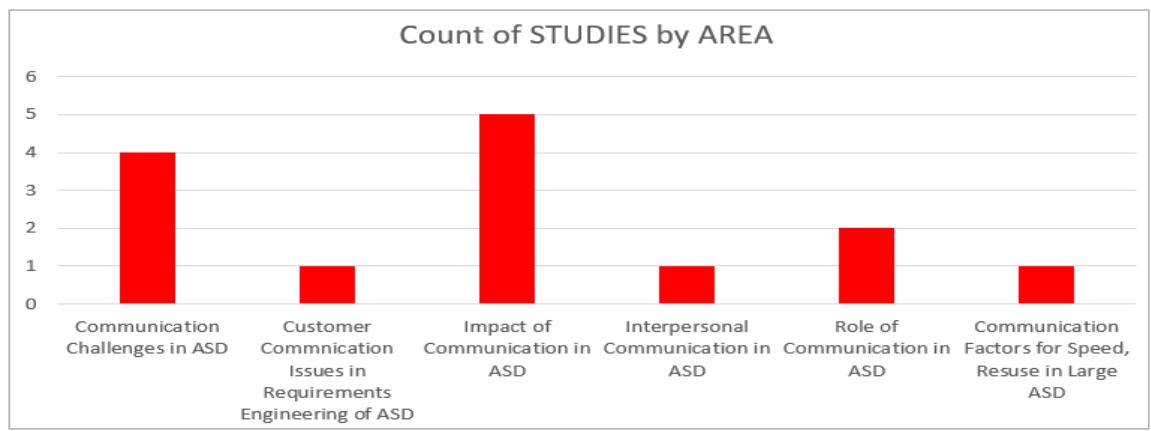

**Fig 2.** Count of research studies by area

*RQ2: What research gaps are there in agile communication?*

The answer to this question will shed light on the potential research avenues that need to be explored. From the answer of our RQ1 we can see that three major areas under communication in ASD have been the focus of the community namely challenges of communication, impact of communication and role of communication in ASD, respectively. However, we believe that there are still multiple avenues of research in this area such as communication between and among ASD teams, communication between all the stakeholders of the project, communication between development, testing and operations team in the context of agile. In our opinion these are the current research gaps that researchers need to address that will help the community move forward.

The results of our RQ1 show that research in ASD communication has mainly focused on the challenges and impact of communication. We see that 42% of the studies worked on the impact of communication and 28% worked on the challenges of ASD. Less that 20% od studies have explored the role of communication. In our opinion this is an underexplored area as communication plays important role in quality and delivery. All the remaining areas contribute approximately 7% of the included studies. RQ2 gives us an idea of where the research is lacking in ASD and we have pointed out those in our answer. We as a community need to actively pursue these research avenues to gather better knowledge on ASD.

**Conclusions and Future Work**

In conclusion we first we would like to point out that ASD has brought a radical change to the software industry and with the growing popularity it is evident that ASD will be future of software industry. The result of this study clearly depicts that multiple of areas of ASD has been explored however there are areas that have not been explored.

From the answer of our RQ2 we can see that communication in ASD has significant research gaps. These gaps are the potential future work in opinion. The community need to investigate these areas to provide better knowledge and guidelines for practitioners. From the existing literature we can see that ASD has been explored. However, ASD at scale has not been explored in detail. We believe all the factors related to ASD will surely change when we especially communication and coordination. Therefore, rigorous primary studies need to be conducted to gather relevant data to move forward.

**Studies identified for this paper**